\documentclass[aip,preprint,cites,nourl,graphicx]{revtex4-1}
\usepackage{epsfig}
\usepackage{booktabs}
\usepackage{amsmath}
\usepackage{amssymb}
\usepackage{color}
\usepackage{setspace}
\usepackage{comment}
\usepackage{subfig}
\usepackage{indentfirst}
\usepackage{bbold}
\begin{document}

\title{A correlated-polaron electronic propagator: open electronic dynamics beyond the Born-Oppenheimer approximation}

\author{John A. Parkhill} 
\email{john.parkhill@gmail.com}
\affiliation{Department of Chemistry and Chemical Biology, Harvard University,12 Oxford St. Cambridge, MA 02138, USA}

\author{Thomas Markovich} 
\affiliation{Department of Chemistry and Chemical Biology, Harvard University,12 Oxford St. Cambridge, MA 02138, USA}

\author{David G. Tempel} 
\affiliation{Department of Chemistry and Chemical Biology, Harvard University,12 Oxford St. Cambridge, MA 02138, USA}

\author{Alan Aspuru-Guzik} 
\email{aspuru@chemistry.harvard.edu}
\affiliation{Department of Chemistry and Chemical Biology, Harvard University, 12 Oxford St. Cambridge, MA 02138, USA}

\begin{abstract}   
\indent 	In this work we develop an approach to treat correlated many-electron dynamics, dressed by the presence of a finite-temperature harmonic bath. Our theory combines a small polaron transformation with the second-order time-convolutionless master equation and includes both electronic and system-bath correlations on equal footing. Our theory is based on the \emph{ab-initio} Hamiltonian, and thus well-defined apart from any phenomenological choice of basis states or electronic system-bath coupling model. The equation-of-motion for the density matrix we derive includes non-Markovian and non-perturbative bath effects and can be used to simulate environmentally broadened electronic spectra and dissipative dynamics, which are subjects of recent interest. The theory also goes beyond the adiabatic Born-Oppenheimer approximation, but with computational cost scaling like the Born-Oppenheimer approach. Example propagations with a developmental code are performed, demonstrating the treatment of electron-correlation in absorption spectra, vibronic structure, and decay in an open system. An untransformed version of the theory is also presented to treat more general baths and larger systems.
\\
\end{abstract}


\maketitle

\section{Introduction}
\indent The small-polaron transformation of the electronic Hamiltonian was originally developed in the 1960's \cite{Devreese:1972qf}, and more recently revived\cite{Dahnovsky:2007vn,Dahnovsky:2007ys} in the many-electron context. It is a classic example\cite{Lang:1962nx} of the utility of canonical transformations in quantum physics. Its usefulness is well-established \cite{Lang:1962nx} yet it is also experiencing renewed interest\cite{Pereverzev:2006fk,Jang:2008ve,Jang:2009kx,Chin:2011fk}. In particular, second-order master equations in the polaron frame afford good results in all bath strength regimes\cite{Nazir:2009uq,McCutcheon:2011ly,Lee:2012fk} employing the variational technique of Harris and Silbey\cite{Silbey:1984fk} . In the many-electron case, there has also been some recent pioneering work toward developing random-phase approximation equations\cite{Dahnovsky:2007vn,Dahnovsky:2007ys}. The electronic structure community has also produced some related work\cite{Monkhorst:1987qf,Micha:1999kl,Dzhioev:2011ly,Krause:2005uq,Tremblay:2010fk,Petrenko:2007hx}, including phenomenologically damped\cite{Kristensen:2011dq} response theory.\\
\indent 	In this paper we formulate and implement a correlated many-electron master equation that overcomes several limitations of the adiabatic Born-Oppenheimer approximation, and includes effects such as excited-state lifetimes and vibronic structure. The basic goal of our method is to produce electronic spectra for small molecules that include the effects of coupling to an environment. With continued development our formalism could describe environmental localization of electronic states and the decay of quantum entanglement in correlated electronic systems. \\
\indent The theory we present exploits the polaron transformation to combine both electron correlation and system-bath couplings in a single perturbation theory. In the transformed frame, high-rank quantum expressions are dressed by environmental factors, which cause them to decay during dynamics. This introduces the possibility for environmentally induced decay of the correlations in an electronic system, thereby making the problem computationally more tractable. We also discuss how a general one-particle perturbation to the electronic system may be treated in a closely related untransformed version of the theory. Physically relevant coupling models of this form are numerous, and several examples include nuclear motion coupled to electronic degrees of freedom \cite{Chapman:2011ve,Parkhill:2012vn}, Coulomb coupling to a nearby nano-particle surface\cite{Morton:2011uq}, the electromagnetic vacuum\cite{Tempel:2011ys}, and perhaps even Coulomb coupling to surrounding molecules in a condensed phase. \\
\indent 	The present method is distinguished from previous work by a few characteristic features. Unlike virtually all master equation approaches, it treats the dynamics \textbf{without assuming the many-body problem of the electronic system to be solved}. We emphasize that our correlated theory is one half of a final theory with the other involving a recipe to calculate system specific couplings with Ehrenfest type dynamics or another scheme. The reason being that, in general, the appropriate basis of a few electronic states to prepare a master equation can't be found \emph{a priori}. As a result, this theory begins from a system-bath Hamiltonian which is well-defined and atomistic in terms of single-electron states and employs a non-Markovian equation of motion (EOM) in place of phenomenological damping. Conveniently, we find that high-rank operator expressions responsible for the computational intractability of exact, \emph{closed}, many-particle quantum mechanics are multiplied by factors which exponentially vanish in many circumstances. \\
\indent 	The picture of electronic dynamics offered by a master equation is complementary to the time-dependent wave-packet picture of absorption\cite{heller:1544,Orel:1980vn,Ben-Nun:1999kx,Kaledin:2003uq,Kay:2005ys,Tatchen:2009kx,Ceotto:2011vn}. In the latter approach the electronic degrees of freedom are described as a superposition of a few adiabatic surfaces, and nuclear wave-packets are studied in the dynamics. In our approach the electronic degrees of freedom are described in all-electron detail, but the dynamics of the nuclei are approximated by the choice of the Holstein Hamiltonian. It is possible to combine a wave-packet calculation of the bath correlation function with the untransformed equation of motion presented in this work, to leverage the physical strengths of both approaches.\\
\indent	We demonstrate the implementation of the formalism in a pilot code, and apply it to calculate some dynamic properties of model small molecules. The spectra produced by the electron correlation theory are shown to compare favorably to related methods in the adiabatic limit. Vibronically progressed spectra are shown to be produced by the dressed theory, and a Markovian version is applied to a basic model of electronic energy transport between chromophores. Finally the undressed, uncorrelated theory is used to simulate the ultraviolet absorption spectrum of 1,1-diflouroethylene and compared with available experimental data. 
\section{Theory}
\subsection{Hamiltonian}
\indent  As in the work of references~\cite{Dahnovsky:2007vn,Dahnovsky:2007ys}, we use the non-relativistic \emph{ab-initio} many-electron Hamiltonian with a Holstein-type\cite{Holstein:1959rq} (linear) coupling to a bath of non-interacting bosons (summation over repeated indices is implied throughout this paper unless stated otherwise), given by 
\begin{align}
\hat{H} = \hat{F} + \hat{V} +\hat{H}_\text{boson}  +\hat{H}_\text{el-boson} =  f_p^qa^\dagger_qa_p + V_{pq}^{rs} a^\dagger_s a^\dagger_r a_p a_q + \omega_k b^\dagger_k b_k + a^\dagger_p a_p M^p_k (b^\dagger_k + b_k)\;. 
\label{eq:ham}
\end{align}
Here, $a^\dagger_s$ creates an electron in the single-particle basis state s, while $b^\dagger_k$ creates a bosonic bath particle in the kth mode. $\hat{F}$ is the Fock operator of the reference determinant, and $\hat{V}$ is the two-electron part of the electronic Hamiltonian in the same single-particle basis. The third term is the boson Hamiltonian, $\hat{H}_b$ and the last is the coupling of the electronic system to the bath. For a general bath mode with dimensionless displacement $Q_k$, the bi-linear coupling constant $M_k^p$, is related to the derivative of the orbital energy via $M_k^p = \omega_k^{-1} \frac{d f_p^p}{dQ_k}$ (no summation over p and q implied). Assuming this sort of coupling is only appropriate for nuclear configurations near the minimum of the Born-Oppenheimer well. There are prescriptions for defining these parameters from normal-mode analysis\cite{Parkhill:2012vn} and molecular dynamics\cite{Gilmore:2008fk} for the semiclassical treatment of anharmonicity. The latter approach can also include electrostatic fluctuations. For the purposes of this work we are only interested in the properties of the method we develop, and do not present any calculations of the bath Hamiltonian. In the following discussion, we will assume the one-electron parts of $\hat{H}$ and $\hat{F}$ to be diagonal in the (canonical) one-electron basis with eigenvalues $\epsilon_p = f_p^p$. 

We now introduce the displacement operator,
\begin{align}
 exp[\hat{S}] = exp[a^\dagger_p a_p \tilde{M}^p_k (b^\dagger_k - b_k)]  \\
\tilde{M}_p^k \equiv M_p^k / \omega_k\;,
\label{eq:transf}
\end{align}
which generates the polaron transformation.
Since the operator $\hat{S}$ is anti-Hermitian, it generates a unitary transformation of the electron-boson Hamiltonian $\hat{H} \rightarrow \tilde{H} = e^{-\hat{S}} \hat{H} e^{\hat{S}}$. Explicitly, the polaron transformed Hamiltonian is given by
\begin{equation}
\tilde{H} = \tilde{F} + \tilde{H}_\text{int},
\end{equation}
where the one-particle part of the transformed Hamiltonian is given by
\begin{align}
\tilde{F} = \tilde{F}_\text{ele} + \tilde{H}_\text{boson} = (\epsilon_p - \lambda_p) a^\dagger_p a_p + \omega_k b^\dagger_k b_k,
\end{align}
and we have introduced the reorganization energy $\lambda_p = \sum_k M^p_k / \omega_k  $. The transformed electron-electron interaction is given by
\begin{equation}
\tilde{H}_\text{int} = \tilde{V}_{pq}^{rs} a^\dagger_s a^\dagger_r a_p a_q X^\dagger_s X^\dagger_r X_p X_q,
\label{eq:int}
\end{equation}
where the transformed matrix elements are
\begin{equation}
\tilde{V}_{pq}^{rs} =  V_{pq}^{rs} - 2(\omega_k \tilde{M}_k^r \tilde{M}_k^s) \delta_{ps} \delta_{qr} (1-\delta_{sr}),
\end{equation}
and we have defined the bath operators $\hat{X}_p$ as
\begin{equation}
\hat{X}_p = exp[\tilde{M}^p_k (b^\dagger_k - b_k)].
\end{equation}
For future reference it will also be useful to define dressed electronic creation and annihilation operators, denoted by $\tilde{a}^\dagger_s \equiv a^\dagger_s X^\dagger_s$ and $\tilde{a}_s \equiv a_s X_s$.

\begin{figure}
\begin{center}
\includegraphics[totalheight=2.1in]{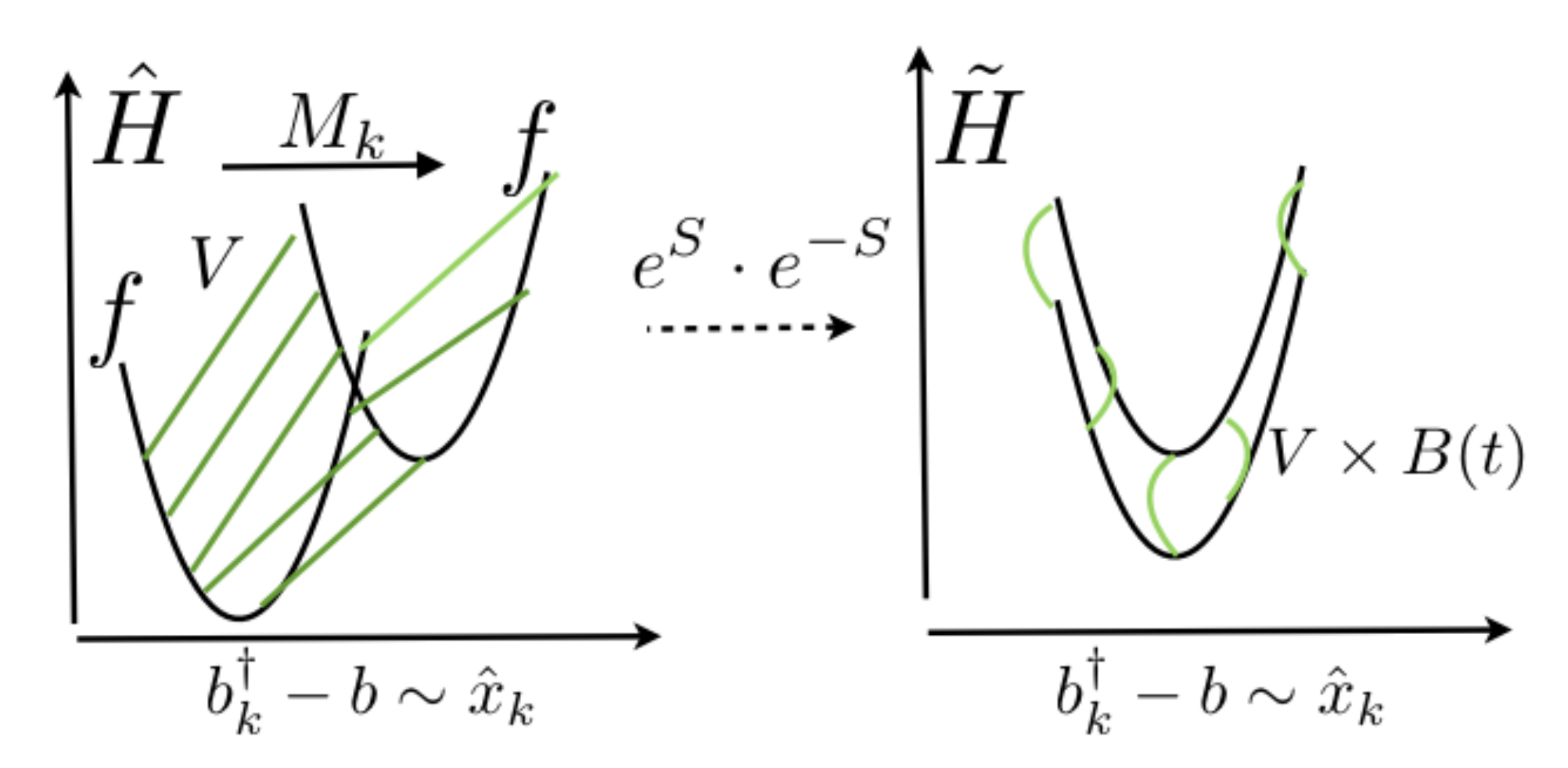}
\caption{A schematic representation of the polaron transformation. One-electron energies of the underlying Hamiltonian take the form of displaced parabolas coupled to the Coulomb interaction $V$. In the transformed frame, the boson effects are absorbed into a $\tilde{V}$, which depends on the boson operators, $\hat{X}$. This makes $\tilde{V}$ effectively time-dependent via the bath correlation function $B(t)$.}\;
\label{fig:scheme}
\end{center}
\end{figure}

\indent The key feature of the polaron transformation is that $\tilde{H}$ has no electron-phonon coupling term.  As a result, the two-electron and electron-boson parts of the original Hamiltonian are combined into a single term which now couples two dressed electrons and two dressed bosons (Eq.~\ref{eq:int}). One should intuitively imagine the situation depicted in Fig. \ref{fig:scheme}, where two displaced electronic energy surfaces are dragged into alignment by the polaron transformation, altering the coupling region between them, which now absorbs the electron-boson coupling. \\
In what follows, our expressions will be derived in the interaction picture with respect to Eq.~\ref{eq:int} and then switched to the Schr\"{o}dinger picture. The harmonic nature of the bosons means that the correlation functions of $\hat{X}$ operators (in any combination and at multiple times) can be given as simple functions of $\omega_p, \tilde{M} \text{ and } \beta=k_b T$\cite{Dahnovsky:2007vn}. The transformed electron-boson problem takes the form of the usual many-electron problem with a time-dependent electron-electron interaction. This allows us to harness powerful methodologies that stem from two distinct areas of research: quantum master equations\cite{Breuer:uq} and quantum chemistry methods\cite{Helgaker:2000ys}. We exploit this feature to produce a model of electronic dynamics which treats system-bath dynamics and correlation effects within the same perturbation theory.\\


\subsection{Electronic dynamics}
\begin{figure}
\begin{center}
\includegraphics[totalheight=1.1in]{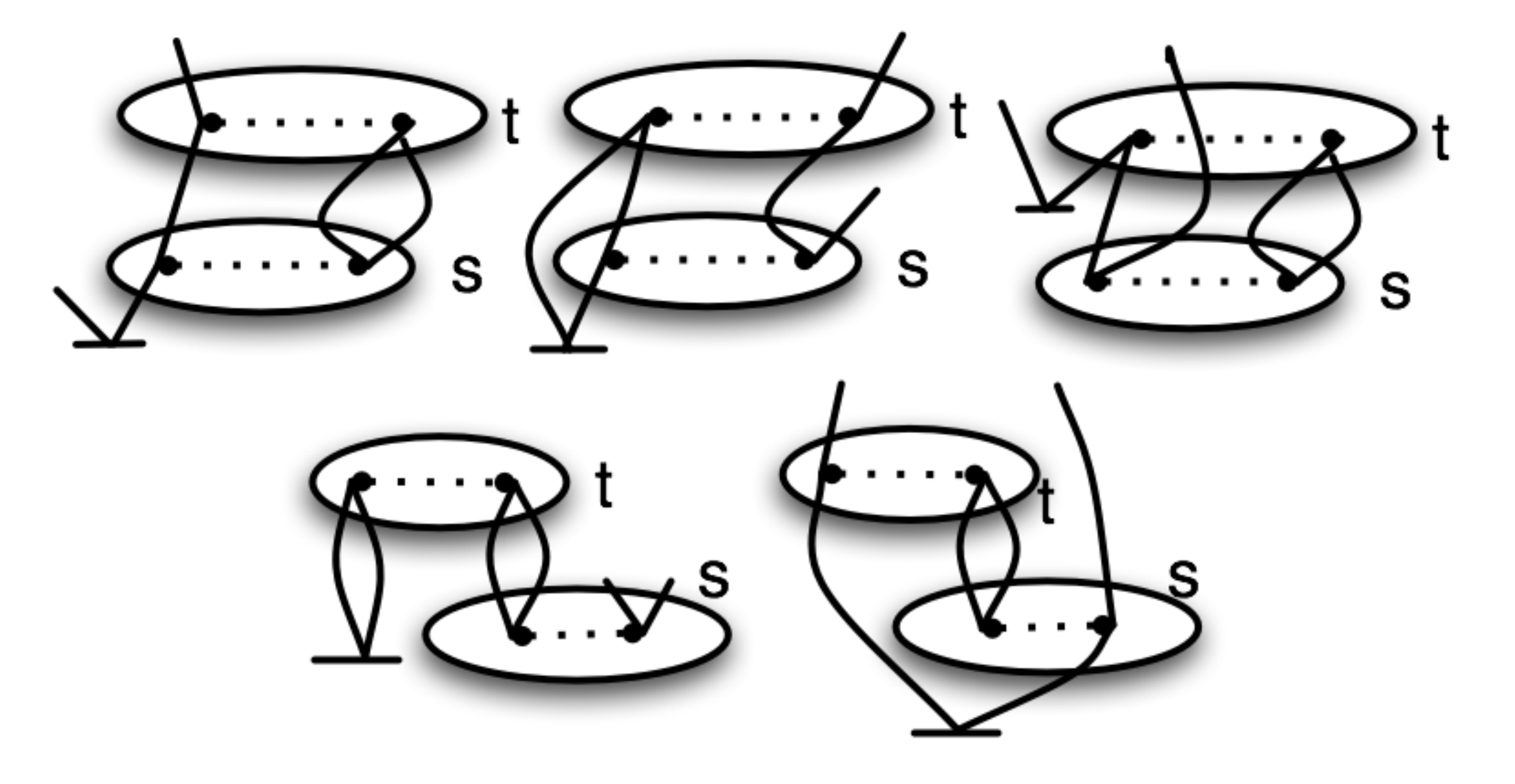}
\caption{ The particle-hole $\rightarrow$ particle-hole skeleton Brandow\cite{BRANDOW:1967kx} diagrams in the second-order homogeneous term, the first two come from the $V_tV_so$ term in the expanded double-commutator and third from $V_t o V_s$. The fourth occurs in $V_t o V_s$, $V_s o V_t$ and $V_t V_s o$. The boson correlation function depends on any lines emerging from the ellipses which represent virtual electron scattering events at times t, s.}
\label{fig:diags}
\end{center}
\end{figure}

\indent 	Having reviewed the polaron transformation in the previous section, we now combine it with the time-convolutionless perturbation theory (TCL) \cite{Shibata:1977tg,Breuer:2001uq} to arrive at the central theoretical result of this manuscript. Our goal is to derive an equation-of-motion for a dressed particle-hole excitation operator, which we denote $\tilde{o}_a^i \equiv o_i^a \tilde{a}^\dagger_a \tilde{a}_i$ (here $i,j,k...$ are zeroth-order occupied levels and $a,b,c... $ are unoccupied). We consider only the particle-hole part of the density matrix, commonly referred to as the Tamm-Dancoff approximation\cite{Dyson:1953dq,Walter:1981ve}(TDA), to simplify the derivation of our equations and avoid the possible issues associated with the non-linearity of the other blocks\cite{Tohyama:1995zr}. To derive Fock-space expressions\cite{Parkhill:2010mw}, it is convenient to assume the initial equilibrium state is the canonical Hartree-Fock (HF) determinant i.e. the initial density matrix is $|\Psi(0) \rangle \langle \Psi(0)| \approx |0\rangle \langle 0|$, with $|0\rangle$ the HF determinant. This assumption relies on two approximations: 1) The first excited-state energy of the systems we will study is much larger than $k_bT$, so that the system is effectively in the ground-state. 2) The initial state of the system is weakly correlated and hence dominated by a single Slater determinant. Assumption 1 is clearly valid for the small molecular systems we will study, while assumption 2 requires more care and the effects of initial correlations are discussed in detail in section II. E.
\indent 	We define projection operators: 
\begin{align}
\mathcal{P} \tilde{\mathcal{A}} = \mathcal{P}_F(Tr_{b}(\tilde{\mathcal{A}})) \times \langle X^\dagger ... X\rangle _{b,eq} \text{, and } \mathcal{Q} = \mathbb{1}-\mathcal{P}. 
\end{align}
$\mathcal{P}_F$ is a Fock space projector onto maps between single-electron $\{a^\dagger a\}$ operators like the typical projector of a partitioned electron-correlation perturbation theory ie: 
\begin{align}
\mathcal{P}_F ( \mathcal{L}_1 : o_{p}^{q} a^\dagger_q a_p \rightarrow \eta_r^s a^\dagger_s a_r ) = \mathcal{L}_1\\\mathcal{P}_F ( \mathcal{L}_{>=2} : o_{pq...}^{...rs} a^\dagger_s a^\dagger_r ... a_p a_q \rightarrow \eta_{p'q'...}^{...r's'} a^\dagger_{s'} a^\dagger_{r'} ... a_{p'} a_{q'} ) = 0.
\end{align}
This partitioning is consistent with the perturbative ordering of $\tilde{H}$ by powers of $\tilde{V}$. We treat the description of the many-electron state in terms of only one-body operators (neglecting all higher density matrices) on the same footing as tracing over the bath degrees of freedom\cite{Alicki:2009vn}, so both phenomena are easily incorporated in the same master equation.\\
\indent The effective Liouvillian becomes time-dependent due to the polaron transformation. Consequently, there exists no simple analytical formula for its Fourier transform. Instead, we must give a differential representation of the EOM for a particle-hole excitation, which can then be integrated numerically. Given these projectors, the time-convolutionless perturbation theory\cite{Shibata:1977tg} (TCL) over the space of $\mathcal{P}$ is: \footnote{Usually the TCL is applied to a phenomenological density operator and 1-body Liouvillian. Here we will apply it to a one-particle transition density operator. The projection operator technique and quantum Ehrenfest theorem required for the TCL both carry over.}.  
\begin{align}
\frac{d}{dt} \mathcal{P}\tilde{o}(t) =  \left( \mathcal{PL}(t)\mathcal{P} + \int_{0}^t ds  \mathcal{PL}(t)\mathcal{Q}\mathcal{L}(s)\mathcal{P} \right ) \tilde{o}(t) + \mathcal{I}(t)
\label{eq:tcl}
\end{align}
This is written in the interaction picture where $\mathcal{L} = -i[\tilde{V}(t),\cdot]$. The first term above is the uncorrelated part of the evolution\footnote{Since the perturbation is a two-particle operator, we cannot trivially diagonalize the first order term, as one does when working with tight-binding type Hamiltonians.}, the second is a homogenous term reflecting correlation between the system and the bath, and the last term is an inhomogeneity reflecting correlations of the initial state. The interaction picture perturbation\footnote{We leave $X$ operators in the interaction picture while pulling electronic operators into the Schodinger picture throughout the text.} is $\tilde{V}(t) = (\tilde{V}_{rs}^{pq} e^{i\Delta_{rs}^{pq}t} a^\dagger_q a^\dagger_p a_r a_s) \times ( X^\dagger_q(t)X^\dagger_p(t)X_r(t)X_s(t))$, where $\Delta_{\alpha \beta...}^{\gamma \delta...} = (\epsilon_\gamma +  \epsilon_\delta + ... - \epsilon_\alpha - \epsilon_\beta - ...  )$. 
Expanding the first term over the one-particle space and moving into the Schrodinger picture we obtain (left arrows, $\leftarrow$, are used to indicate the contribution of a  term to ${d \tilde{o}}/{dt}$ ): 
\begin{align}
\frac{d}{dt} \mathcal{P}\tilde{o}_v^u(t) \leftarrow -i(\epsilon_u - \epsilon_v+   \tilde{V}_{us}^{vp}B_{pu}^{vs}(t,t) ) \tilde{o}_p^s (t) 
\label{eq:Term1}
\end{align}
\indent 	We have introduced a shorthand for the correlation function $B_{(a^\dagger),(a)}^{(a^\dagger),(a)}($lower time, upper time):
\begin{align}
B_{p q r s}^{m n o p} (t,s) = \langle X_m^\dagger(s)  X_n^\dagger(s)  X_o(s)  X_p (s) X_p^\dagger(t)  X_q^\dagger(t)  X_r(t)  X_s (t). \rangle 
\end{align}
Term (\ref{eq:Term1}) is comparable in dimension and physical content to the response matrices of configuration interaction singles (CIS)\cite{Greenman:2010zr} with an attached, time-local boson correlation function. Because all arguments have the same time-index, $B$ only applies a real factor ($\sim 1 \text{ as } \tilde{M}\rightarrow 0$) to values of the interaction and thus introduces no new time-dependence. The indices of the boson correlation function, and their order, are simply read-off the $\tilde{V}$ integral they multiply. \\
\indent 	The second homogeneous term introduces bath correlation functions between boson operators occurring at different times (t and s) according to
\begin{align}
\frac{d}{dt} \mathcal{P}\tilde{o}_v^u(t) \leftarrow \left (\frac{-i}{\hbar} \right)^2 \int_{t_0}^t ds (  \mathcal{PL}(t)\mathcal{QL}(s)\mathcal{P} \tilde{o}(t)) \\
 = - [\tilde{V}_{pq}^{rs}(t), \mathcal{Q} [ \int_{t_0}^t B_{rs,qp}^{ab,xy}(t,s) \tilde{V}_{xy}^{ab}(s) ds ,\tilde{o}_m^n(t)]] 
\label{eq:Term2}
\end{align}
Moving the electronic part into the Schodinger picture and rearranging this becomes:
\begin{align}
 = - [\tilde{V}_{pq}^{rs},\mathcal{Q} [ \tilde{V}_{xy}^{ab} ,\tilde{o}_m^n ]]  e^{i (\Delta_{pq}^{rs}) t} e^{i (\Delta_{m}^{n}) t} \int_{t_0}^t B_{rs,qp}^{ab,xy}(t,s) e^{i (\Delta_{xy}^{ab}) s} ds
\label{eq:Term2p}
\end{align}
Expanding the commutators in Eq. \ref{eq:Term2p}, applying Wick's theorem to remove the many vanishing terms, and enforcing the connectivity constraint, one obtains many topologically distinct terms. In our implementation this is done automatically before the execution of a simulation. The terms are easily related to terms which occur in the expansion of the second-order Fermion propagator (SOPPA)\cite{Alberico:1991bh,Simons:1973uq,Jorgensen:1975fk} and diagonalization-based excited-state theories like CIS(D)\cite{Hirata:2005oq}. Since we employ the time-convolutionless perurbation theory, the oscillating $e^{i\Delta t}$ factors are different than those which occur in Rayleigh-Schrodinger perturbation theories (in the energy-domain denominator $\frac{1}{\omega - \Delta}$) and warrants further study. We some terms from a hole $\rightarrow$ particle excitation which is obtained by multiplying both sides of equation (\ref{eq:tcl}) on the left with $(a^\dagger_u a_v)^\dagger$ and applying Wick's theorem. 
\begin{align}
  \dot{\tilde{o}}_i^a (t) \leftarrow \tilde{V}_{bd}^{aj} \tilde{V}_{cj}^{bd} \tilde{o}_i^c  e^{i (\Delta_{bd}^{cj}) t} \int_{t_0}^t B_{aj,bd}^{bd,cj}(t,s) e^{i (\Delta_{cj}^{bd}) s} ds
\label{eq:Term2pp}
\end{align}
This term has six indices overall, but can be factorized as follows: 
\begin{align}
  \dot{\tilde{o}}_i^a (t) \leftarrow I_c^a(t) \tilde{o}_i^c  \text{ where: } I_c^a(t) = \tilde{V}_{bd}^{aj} \tilde{V}_{cj}^{bd} e^{i (\Delta_{bd}^{cj}) t} \int_{t_0}^t B_{aj,bd}^{bd,cj}(t,s) e^{i (\Delta_{cj}^{bd}) s} ds
\label{eq:Term2ppp}
\end{align}
The calculation of $I_c^a$ scales fifth order with the number of single electron states, and linearly with the number of bath modes (which go into the calculation of $B(t)$). This low scaling with number of bath modes is inherited from the approach of Silbey\cite{Silbey:1984fk}, Jang\cite{Jang:2009kx} and Nazir\cite{Kolli:2011zr}. The algebraic version of the second term in Fig. \ref{fig:diags} is:
\begin{align}
 \dot{\tilde{o}}_k^b (t) \leftarrow \tilde{V}_{ak}^{ij} \tilde{V}_{cj}^{ab} \tilde{o}_i^c  e^{i (\Delta_{ab}^{cj}) t} \int_{t_0}^t B_{ij,ak}^{ab,cj}(t,s) e^{i (\Delta_{cj}^{ab}) s} ds
\label{eq:TermBad}
\end{align}
This term is sixth order with the size of the system, with the boson correlation function preventing a desirable factorization of the V-V contractions. 14 sixth-or-less-order terms are found which couple hole-particle excitations to each other; skeletons\cite{BRANDOW:1967kx} of these are shown diagrammatically in Fig. (\ref{fig:diags}) with explicit expressions for each of these terms given in the supplementary information\cite{supplementaryInfo}.
We should note that our formalism lacks several terms which occur in the SOPPA because of the $\mathcal{Q}$ projector and the absence of $V_t o_\text{vac} o V_s$ ordered terms that are invoked in the formal expansion of the time-ordered exponential.\\
\indent 	As described so-far the EOM is not time-reversible even when $\tilde{M} = 0$ because neither the homogenous term nor the normal excitation operator take the form of an anti-Hermitian matrix.\footnote{This same issue occurs in SOPPA and CIS(D). Since CIS(D) is derived from diagonalization of the response matrix, this problem appears as a non-Hermitian effective response matrix instead of time-irreversibility, but these issues are the same.}. We remedy this by adding the terms which result from the normal excitation's operator's Hermitian conjugate, ie: the indices of the vacuum and $o$ are swapped, and the signs of $\Delta_s$ and prefactor are flipped and the new term is added to the other perturbative terms. For example Term \ref{eq:TermBad} becomes the following two terms:  
\begin{align}
 \dot{\tilde{o}}_k^b \leftarrow \frac{1}{2}\tilde{V}_{ak}^{ij} \tilde{V}_{cj}^{ab} \tilde{o}_i^c \int_{t_0}^t B_{ij,ak}^{ab,cj}(t,s) ( e^{i (\Delta_{ab}^{cj}) (t-s)})ds  \notag \\
   \dot{\tilde{o}}_i^c \leftarrow \frac{-1}{2}\tilde{V}_{ak}^{ij} \tilde{V}_{cj}^{ab} \tilde{o}_k^b \int_{t_0}^t B_{ij,ak}^{ab,cj}(t,s) ( e^{i (\Delta_{cj}^{ab}) (t-s)})ds  
 \end{align}
Making this modification, the linear response spectrum of the adiabatic model is not significantly altered, but the adiabatic norm conservation is enforced, and changes by less than 1 percent after 1700 atomic units in the case of $H_4$ with 4$^{th}$ order Runge-Kutta and a timestep of 0.05 au.\\

\subsection{Dipole Correlation function}
\indent 	Like any canonical transformation\cite{Yanai:2007fj}, the polaron transformation preserves the spectrum of the overall electron-phonon Hamiltonian but the statistical meaning of the state related to a particular eigenvalue is changed. In other words: the operators of our theory are different objects from the adiabatic Fock space. To lowest order in system-bath coupling\cite{Kolli:2011zr}, electronic observables like the time-dependent dipole correlation function, which predicts the results linear optical experiments, can be obtained by generating the transformed property operator $\tilde{\mu} = e^{\hat{S}} \hat\mu e^{-\hat{S}} = \mu^i_j a^\dagger_i a_j X^\dagger_iX_j$. Within the Condon approximation the dipole correlation function is then the product of the electronic trace and bath trace over the $\hat{X}$ operators introduced in $\tilde{\mu}$. This adds a bath correlation function to the usual observable. Taking all the relevant expectation values gives the following explicit numerical formula for the dipole moment expression:
\begin{align}
C_{d-d}(t) = \sum_{ijab} \{\mu_{ia} \tilde{o}_{ia}(t) \mu_{jb} \tilde{o}_{jb}(0) \} \cdot Tr_B\{ X_a^\dagger(t) X_i (t) X_b^\dagger(0) X_j(0)) \}
\end{align}
This expression for the dipole-dipole correlation function assumes the bath remains at equilibrium. CIS likewise only offers a zeroth-order oscillator strength\cite{Schirmer:1982fu}. The spectra in this work are generated by ``kicking'' the electronic system with the dipole operator instantaneously, and Fourier transforming the resulting dipole-dipole oscillations. 
\subsection{Untransformed Version}
\indent 	Because of the polaron transformation the above formalism is accurate in the strong bath regime with diagonal system-bath couplings. To treat an off-diagonal, weak coupling one can develop the complementary untransformed theory. With a bi-linear system-bath coupling of the form:
\begin{align}
H_{sb} = \sum_{ij} a^\dagger_i a_j M^{ij}_k \left ( b^\dagger_k + b_k \right),
\end{align}
development of an uncorrelated particle-hole equation of motion follow similarly to the above with $H_{sb}$ taking the place of $\tilde{V}$. The untransformed version also makes it possible to introduce an Ehrenfest scheme for the nuclear bath which may be pursued in future work. The projector of the untransformed version is simply the equilibrium trace over $b$ operators. The TCL produces second-order contribution of the form: 
\begin{align}
\dot{\hat{o}}(t) \leftarrow \int^{t}_{t_{0}} [H_{sb}(t),[H_{sb}(s),\hat{o}(t)]] ds
\end{align}
which after translation into the Schrodinger picture and application of Wick's theorem produces several terms\footnote{Given in the supplementary material\cite{supplementaryInfo}.} similar to the following: 
\begin{align}
\dot{o}^a_l (t) \leftarrow  M_{k}^{j, l} M_{k}^{m, j} o^a_m(t)  e^{i\Delta^{j}_{k}t} \int_{t_0}^t  e^{i\Delta_j^m s}C(t-s) ds  
\end{align}
with $\tau$ $ = $ $t-s$. The correlation function $C_m(t) = \langle b_m^\dagger(t) b_m(0)\rangle$ is given by the usual formula\cite{Mukamel:1995zt}:
\begin{align}
\mathfrak{Re}\left(C(\tau) \right) &= \sum_{i} \coth \left ( \frac{ \beta \omega_i}{2} \right ) \cos \left( \omega_i z \right) \notag \\
\mathfrak{Im}\left(C(\tau) \right) &= \sum_{i} \sin \left( \omega_i z \right).
\end{align}
These contributions to the equations of motion scale with the 4$^{th}$ order of the system size; two-orders of magnitude cheaper than the transformed version. They can be combined with just the uncorrelated part of the particle-hole equation of motion, or added to the correlation terms developed above with the bath factor removed. To treat a continuous bath of oscillators, one can introduce a continuous parameterization of $(M_k)^2 = J(\omega)/\omega^2 = {\eta_i} \frac{\omega} e^{-\omega/\omega_c}$ as an ohmic spectral density under the assumption that the coupling to the continuous bath is the same for every state up to a factor. 

Furthermore, we find that the reorganization energy is given by $\lambda_i = \omega_i d^2_{ii}$ which can be subtracted from the original Hamiltonian.

\subsection{Initial Correlations}
\indent	In the above we have expanded on existing master equations by giving the electronic system a many-body Hamiltonian with Fermionic statistics. We have also improved typical electronic response theories by incorporating bath dynamics, but we have assumed that the initial state was a single determinant \emph{and} employed the corresponding Wick's theorem. These approximations are made in many other treatments of electronic spectra\cite{Christiansen:1995kl,Head-Gordon:1999qf,Chapman:2011ve}, and are acceptable for situations where a molecule begins in a nearly determinantal state. Optical absorption experiments of gapped small molecules belong to this regime. To rigorously study electron transport in a biased junction\cite{Myohanen:2008vn}, or otherwise more exotic initial state requires a treatment of initial correlations\cite{Hall:1975ly,Danielewicz:1984kx,BONITZ:ve,Dahlen:2007ys}, as in the non-equilibrium Green's function(NEGF) method\cite{Morozov:1999ve,Garny:2009uq,Leeuwen:2012fk}. These methods first propagate an initial determinant in imaginary time \cite{Myohanen:2008vn} thermalizing and correlating the system before the dynamics, but numerical applications of the NEGF formalism require storage an manipulation of a state variable with three indices: $G_{pq}(\omega)$, and are usually limited to small systems\cite{Balzer:2010qf}, and short times. We can similarly perform an imaginary time propagation, and have performed some exploratory calculations which are described in the supporting supporting material\cite{supplementaryInfo}.\\
\indent 	Instead of performing an imaginary time integration, one can replace the usual Wick's theorem, and build the theory beginning from a state which is already correlated. Extended normal ordering\cite{Kutzelnigg:1997rb}, makes it possible to take expectation values with a multi-configurational reference function via a generalization of Wick's theorem\footnote{a related technique was recently develop for the Green's function\cite{Leeuwen:2012fk}}. Using the extended normal ordering technique, the expressions of the present paper can be promoted to treat a general initial state directly, so long as the density operator of that state is known. This approach would eliminate the need for any adiabatic preparation of the initial state, but would generate more complex equations, and is a matter for future work.\\
\indent 	For weakly correlated systems it's reasonable to keep the approximation that the usual Wick's theorem applies to the initial state of the electronic system, and instead of adiabatically preparing an initial state, choose a perturbative approximation to the correlated part of the initial state, $\mathcal{Q}\tilde{o}(0) = \int_{-\infty}^{0}ds \mathcal{L}(s)\tilde{o}(0)$, and treat the first inhomogenous\cite{Uchiyama:2009kx} term of the master equation: 
\begin{align}
 \mathcal{I}(t) = \mathcal{P}\mathcal{L}(t) \mathcal{Q} \int_{-\infty}^{0}ds \mathcal{L}(s)\tilde{o}(0)
\label{eq:tcl}
\end{align}
The bath parts of the inhomogenous term are known to be relatively unimportant from studies of tight-binding Hamiltonians\cite{Jang:2009kx}, and only slightly perturb the results of a propagation for short times, and so we will not include them.\\ 
\indent 	It is interesting to note that with the addition of this inhomogenous term, there is a correspondence between the present theory and a model of electronic linear response derived from an effective Hamiltonian, CIS(D)\cite{Head-Gordon:1999qf}. The CIS(D) excited states are the eigenvectors of a frequency dependent matrix, $A_{ai,bj}^{CIS(D)}(\omega)$, with the dimension of the particle-hole space: 
\begin{align}
	A_{ai,bj}^{CIS(D)}(\omega) = \hat{H}^{CIS} - \frac{\mathcal{P}\hat{V}\mathcal{Q}\hat{V}\mathcal{P}}{(\Delta - \omega)} + \mathcal{P}\hat{V}\mathcal{Q} \hat{T}_2 \mathcal{P}
\end{align}
In the above $\hat{T}_2$ is the second-order excitation operator from Moller-Plesset perturbation theory. These terms correspond respectively to the Fourier transform of our $\mathcal{PVP}$, $\mathcal{PVQVP}$, and $\mathcal{I}(t)$ terms with different denominators. This correspondence suggests the present method should produce linear response spectra of quality similar to CIS(D), which is usually slightly better than TDDFT. Because the formalism is somewhat involved and many approximations have been made, we have summarized the limitations of this work in a table (\ref{tbl:appx}) with references that point to possible improvements. The formalism is now developed to the point where particle-hole excitations can be usefully propagated, and the main features of the approach can be demonstrated in calculations. 
\begin{table}
\begin{tabular}{ l  r }
\toprule
\hline
Approximation & Extension \\
\hline
\midrule
  $|\Psi_{eq}\rangle \approx |0\rangle$ & Extended normal ordering\cite{Kutzelnigg:1997rb} \\
  Second-Order in $\tilde{V}$ & TCL-4\cite{Timm:2011bs} \\
   $\langle \mu (t) \rangle \approx Tr(\tilde{\mu} \cdot \tilde{o}(t))$  & Use basis commuting with $\mu$ \\
  TDA & Include the additional blocks\cite{Schirmer:1982fu}.\\
  Orbital relaxation & Variational condition\cite{Silbey:1984fk} \\ 
\hline
  \bottomrule
\end{tabular}
  \caption{Limitations of this work and how they may be relaxed. Orbital relaxation isn't really an approximation, per-se, but would be beneficial for the results of the perturbation theory.}
  \label{tbl:appx}
\end{table}
\section{Results}
\subsection{Adiabatic ($\tilde{M} \rightarrow 0$) spectrum}
\indent 	To incorporate both bath and electron correlation effects it was necessary to write down a second-order, time-local EOM for electronic dynamics based on the time-convolutionless perturbation theory which we will call "2-TCL". The zeroth order poles of the correlation terms in this theory differ from those which occur in other second-order theories of electronic response (SOPPA\cite{Oddershede:1977vn}, ADC(2)\cite{Schirmer:1982fu} , CIS(D)\cite{Head-Gordon:1999qf}, and CC2\cite{Christiansen:1995kl}) which arise from perturbative partitioning of what is essentially an energy-domain propagator matrix. Interestingly, the denominator of the present theory is naturally factorized and in the adiabatic limit all terms can be evaluated in fifth order time, unlike Rayleigh-Schodinger perturbation theory which requires a denominator factorization approximation to avoid a 6-index denominator\cite{Hirata:2005oq}. To verify that the electronic part of this work is indeed a reasonable model of electronic dynamics and check our implementation (signs, factors etc.) it is useful to compare an adiabatic spectrum (a calculation with no bath coupling, $\tilde{M}\rightarrow 0$) to one arising from exact diagonalization. We've coded the above formalism into a standalone extension of the Q-Chem package\cite{Shao:2006ys} from which we take the results of some other standard models. The particle hole equation of motion is integrated with the Runge-Kutta $4^{th}/5^{th}$ method with an adaptive time step. Propagations in all three directions are followed at once, and the resulting time dependent dipole tensor is Fourier transformed to produce the spectra presented below. Bath integrals are calculated with 3rd-order Gaussian quadrature with the same time-step as the electronic Runge-Kutta integration, or integrated analytically in the adiabatic limit. The exact results and moments shown below come from the PSI3\cite{Crawford:2007zr} program package. \\
\indent 	To check the adiabatic theory, we present calculations of dipole spectra on the $H_4$ and BH$_3$ molecules\footnote{Coordinates: ((Bohr) H: -1, 0, 0. ; H: 1, 0, 0 ; H: -2.17557, 1.61803, 0; H: 2.17557, 1.61803, 0) and (B (\AA): -0.26429,0.47149,0 H: 0.84371, 0.47149, -0.40000; H: -0.81829, 1.43104, 0.4; H: -0.81829, -0.48807,0.4)}. In both cases the molecules have been stretched from their equilibrium bond lengths to a geometry where correlation effects are stronger\cite{Casanova:2009ly} and excitations are anomalously low-in-energy because of near degeneracy of the single-particle levels. We prepared the minimal basis molecule of H$_4$ in a density excited by the dipole operator, and propagated it for 250au. The dipole-dipole correlation functions $C_{\alpha\beta}(t) = \langle \mu_\alpha(t) \mu_\beta(0) \rangle $ were collected during the simulation and Fourier transformed. The real part of this spectrum of spherically-averaged dipole oscillations is compared against stick spectra with the height given by the transition moment at the poles of an exact \emph{adiabatic} calculation within this basis, and related theories. 
\begin{figure}
\begin{center}$
\begin{array}{cc}
\includegraphics[totalheight=2.3in]{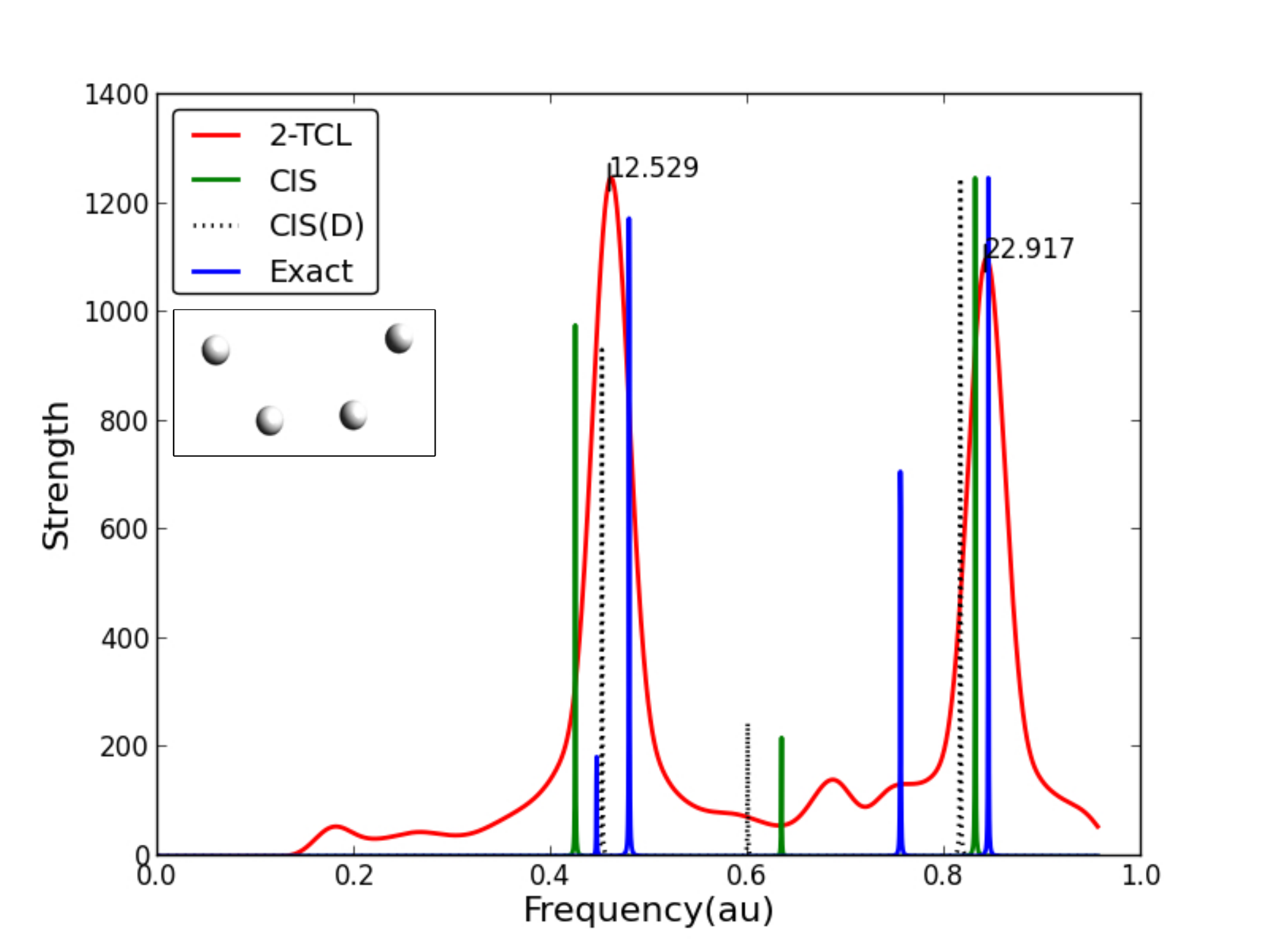} & 
\includegraphics[totalheight=2.3in]{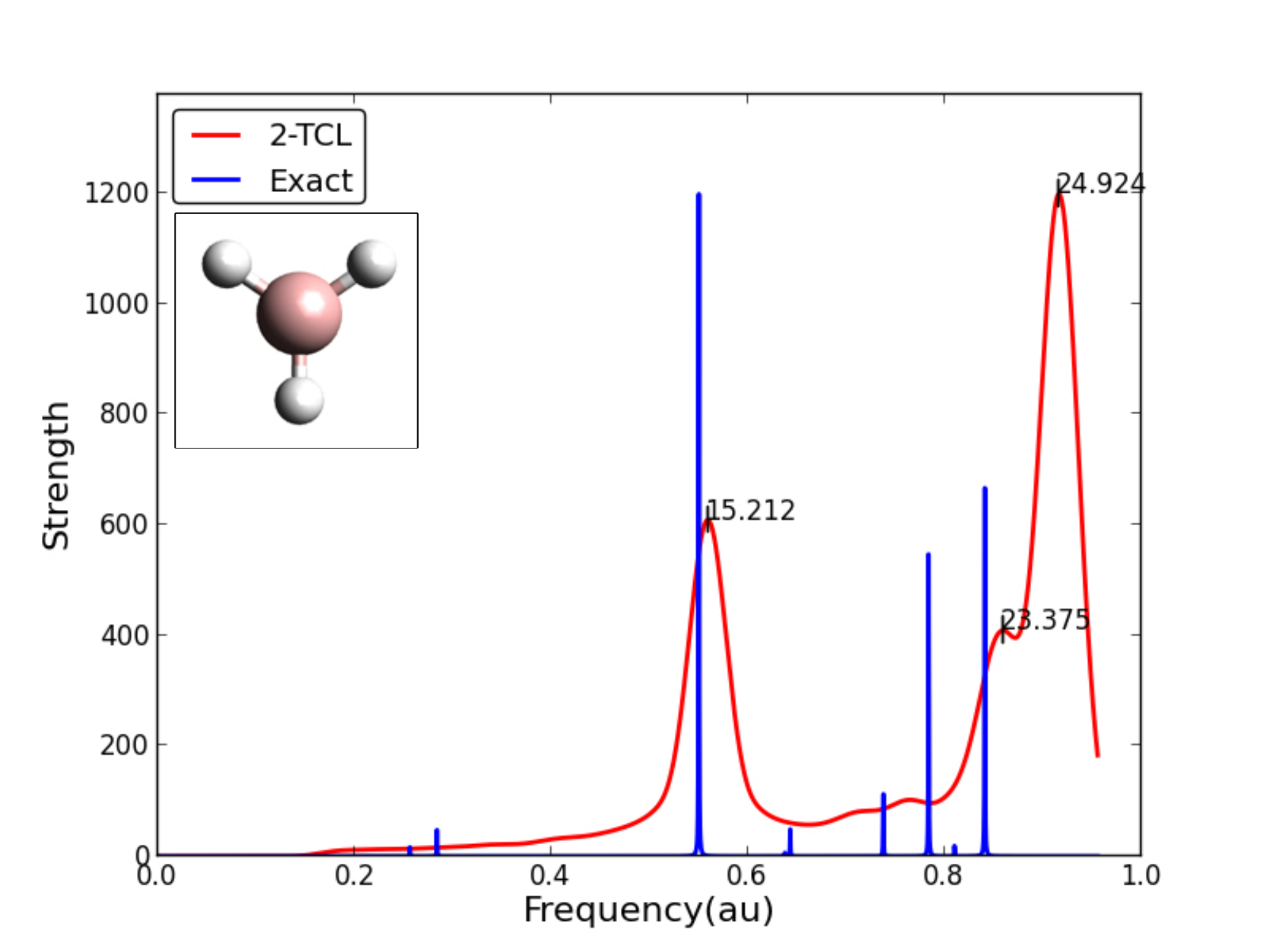}
\end{array}$
\caption{ Left: Adiabatic dipole absorption spectra of H$_4$ in the minimal basis, CIS which is a part of the first term in \ref{eq:tcl}, CIS(D), and the present theory. Numbers indicate maxima in eV and horizontal axis is given in Hartree. The largest maxima have been normalized to the same value. Right: adiabatic 2-TCL applied to the BH$_3$ molecule. The placement of red peaks nearer to blue than green indicates success as a perturbation theory. The atomic geometries (given in a footnote, are also shown).}
\label{fig:AdiaSpectra}
\end{center}
\end{figure}
One sees that the propagator of the present work is a meaningful correction to CIS, moving poles away from the green positions towards the blue (Fig. \ref{fig:AdiaSpectra}) as CIS involves exact diagonalization of the single-excitation part of the many electron Hamiltonian while the present theory involves correlation effects as well. In the adiabatic limit of the response model, only correlation effects (and not bath effects) are present. The excited state near 12.52eV in H$_4$ has increased in energy from the green position of 11.54eV toward the exact pole at 13eV. A more strongly correlated state just below this is unresolved. This is likely because the dipole moment operator doesn't couple the reference determinant to this multi-configurational state. The higher energy peak of the spectrum is brought into good agreement with the exact result.  Similar performance is seen in the case of a distorted BH$_3$ molecule, albeit with corrections of the higher energy peaks near 23eV being too small. \\

\subsection{Vibronic features}
\begin{figure}
\begin{center}
\includegraphics[totalheight=2.4in]{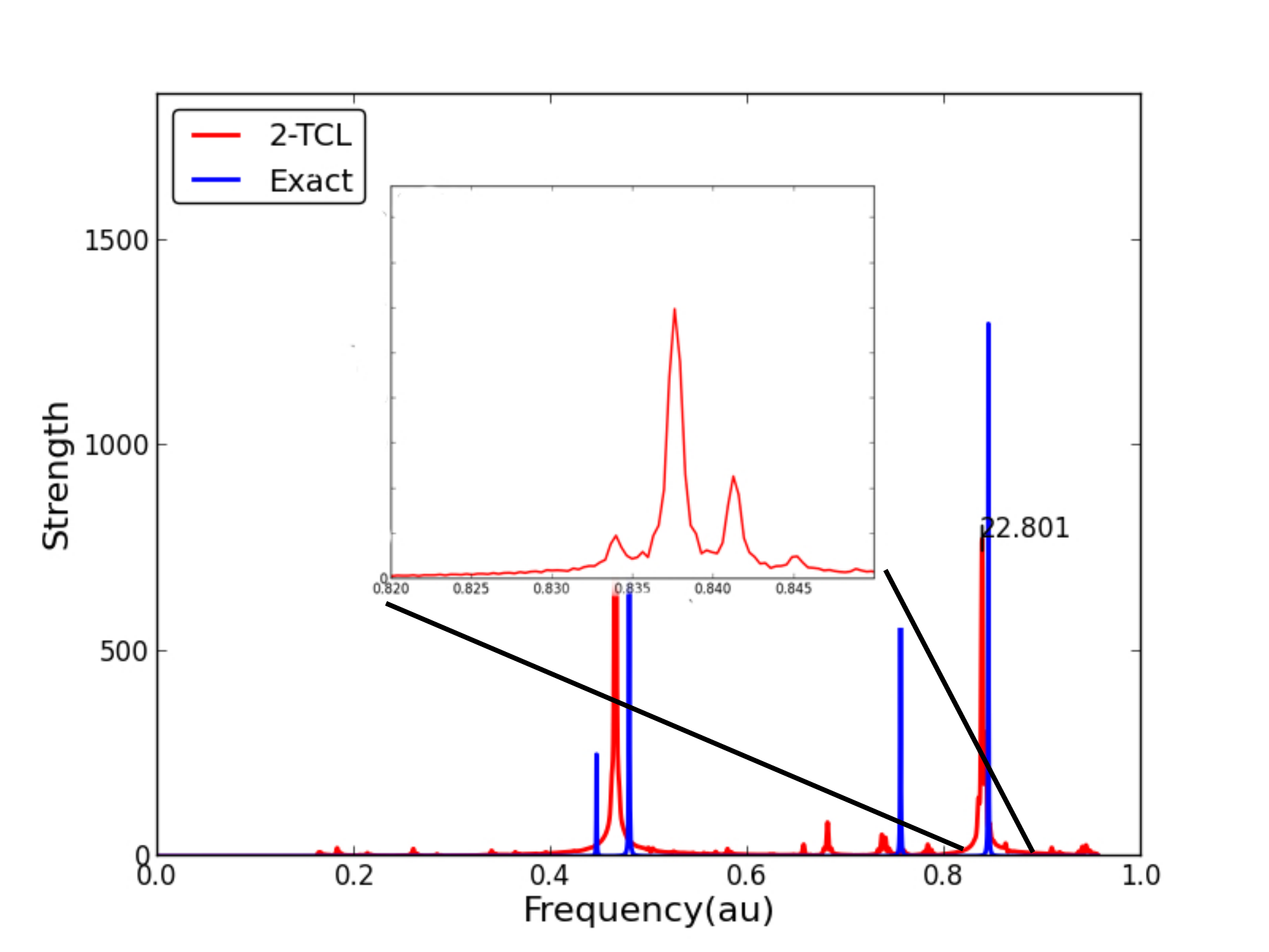}
\caption{ \emph{Polaron-transformed} absorption spectrum of $H_4$ with a Lorentizan bath oscillator at 1600cm$^{-1}$. Each adiabatic peak is split into a Boltzman-weighted vibronic progression magnified at inset. }
\label{fig:sideband}
\end{center}
\end{figure}

\indent 	The spectra of Markovian system-bath perturbation theories take the form of Lorentzians\cite{Tempel:2011ys} at the poles of the response matrix. 
 Because the present theory is non-Markovian it should be capable of yielding new poles off main transitions. To evaluate this effect we add a strong bath oscillator at ~1600cm$^{-1}$ and calculate absorption spectra. We choose this bath oscillator at high energy to observe vibronic peaks in a reasonable amount of propagation time (1700 a.u.), and choose a correspondingly high temperature (4397.25 K) so that there are several peaks in the progression which should take-on a Boltzmann-ian shape. The resulting absorption spectrum is pictured in Figure \ref{fig:sideband}, with a close up of the promised progression. 
%
%
It is relevant to wonder whether the vibronic peaks are simply the result of the bath displacement operators present in the dipole correlation function expression, or the result of the polaron density matrix dynamics. Simply eliminating the bath-correlation function from the dipole expression and generating the same spectrum, a vibronic progression still results, indicating that it is the non-Markovian dynamics of the system operators which provides the vibronic progression. 

\subsection{Energy Transport and Markovian Evolution}
\begin{figure}
\begin{center}
\includegraphics[totalheight=2.1in]{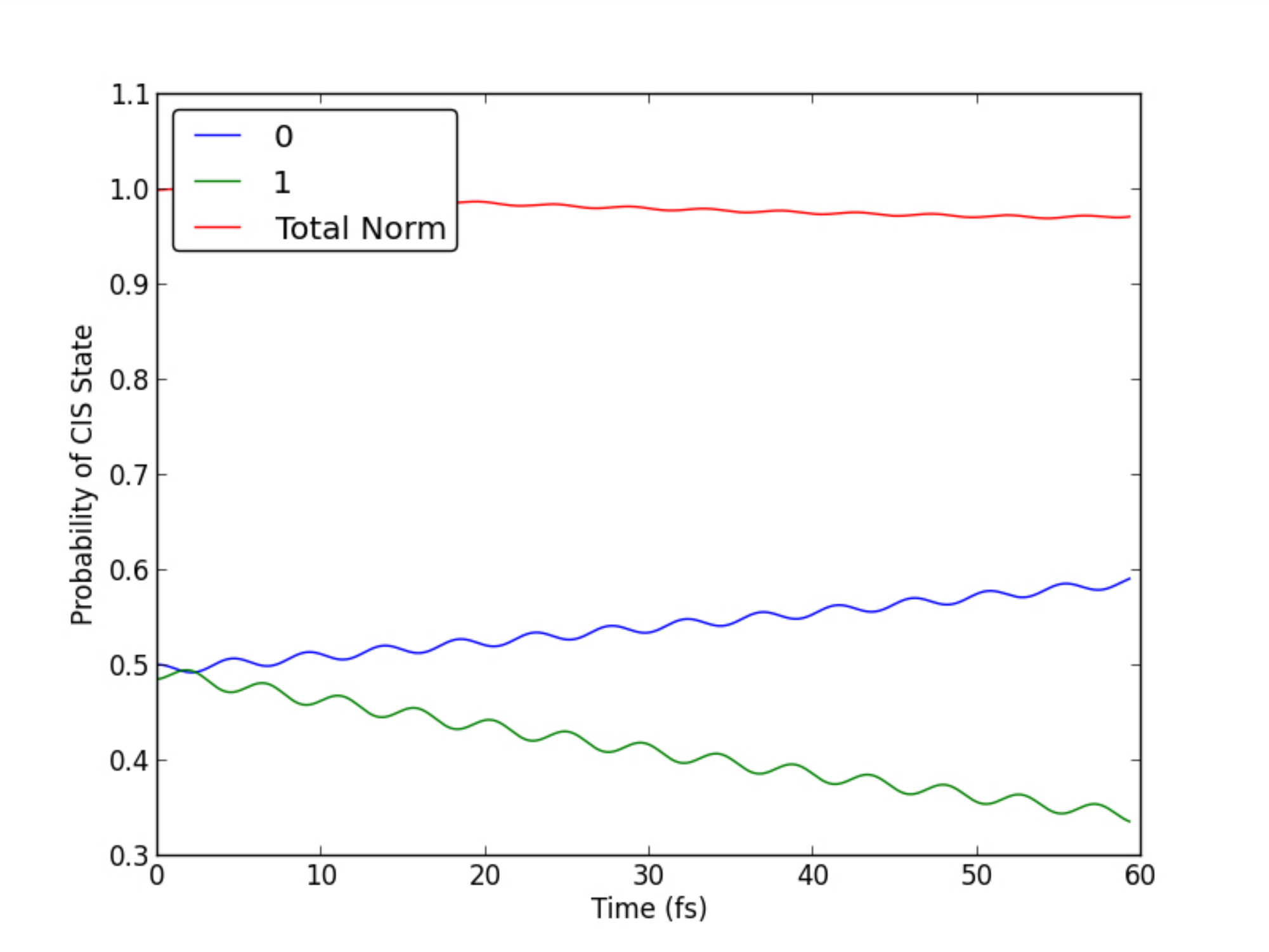}
\caption{Transfer of probability between  H$_2$ molecules in the DZ basis. After being initialized in a transition density which is the superposition of the adiabatic states at 17.3 and 18.1 eV, probabilities relaxation proceeds because of interaction with the bath. To zeroth-order in bath coupling State 1 is localized on the left, undistorted H$_2$, and State 0 is localized on the right, distorted H$_2$.}
\label{fig:Trans}
\end{center}
\end{figure}

\indent 	The continuous integration of rank-6 bath correlation tensors required for the Non-Markovian propagation executed above is a rather costly proposition for large systems. This is especially true if one would like to study incoherent electronic energy transport which takes place in times on the order of picoseconds, roughly a million times more than the electronic timestep required to integrate Eq. (\ref{eq:tcl}) for a typical molecule. A useful approximation to overcome this is a Markov approximation, by which we mean taking the limit of the integral to infinity for each term, such as: 
\begin{align}
R_{ij,ak}^{ab,cj} = \lim_{t \to \infty} \int_{t_0}^t B_{ij,ak}^{ab,cj}(t,s) ( e^{i (\Delta_{ab}^{cj}) (t-s)})ds  
\label{eq:Markov}
\end{align}
where $R$ is now a factor replacing the integral expression. Each term possesses its own $R$ tensor, but these must only be calculated once. Numerically, this limit can be taken in the case of a continuous super-ohmic bath with cutoff frequency $\omega_c$ by introducing a cutoff time $t_c$, above which the bath correlation function is assumed to be equal to it's equilibrium value, which is a good approximation for $\beta\omega_c(t_c)^2 >> 2$.\\
\begin{figure}
\begin{center}
\includegraphics[totalheight=2.1in]{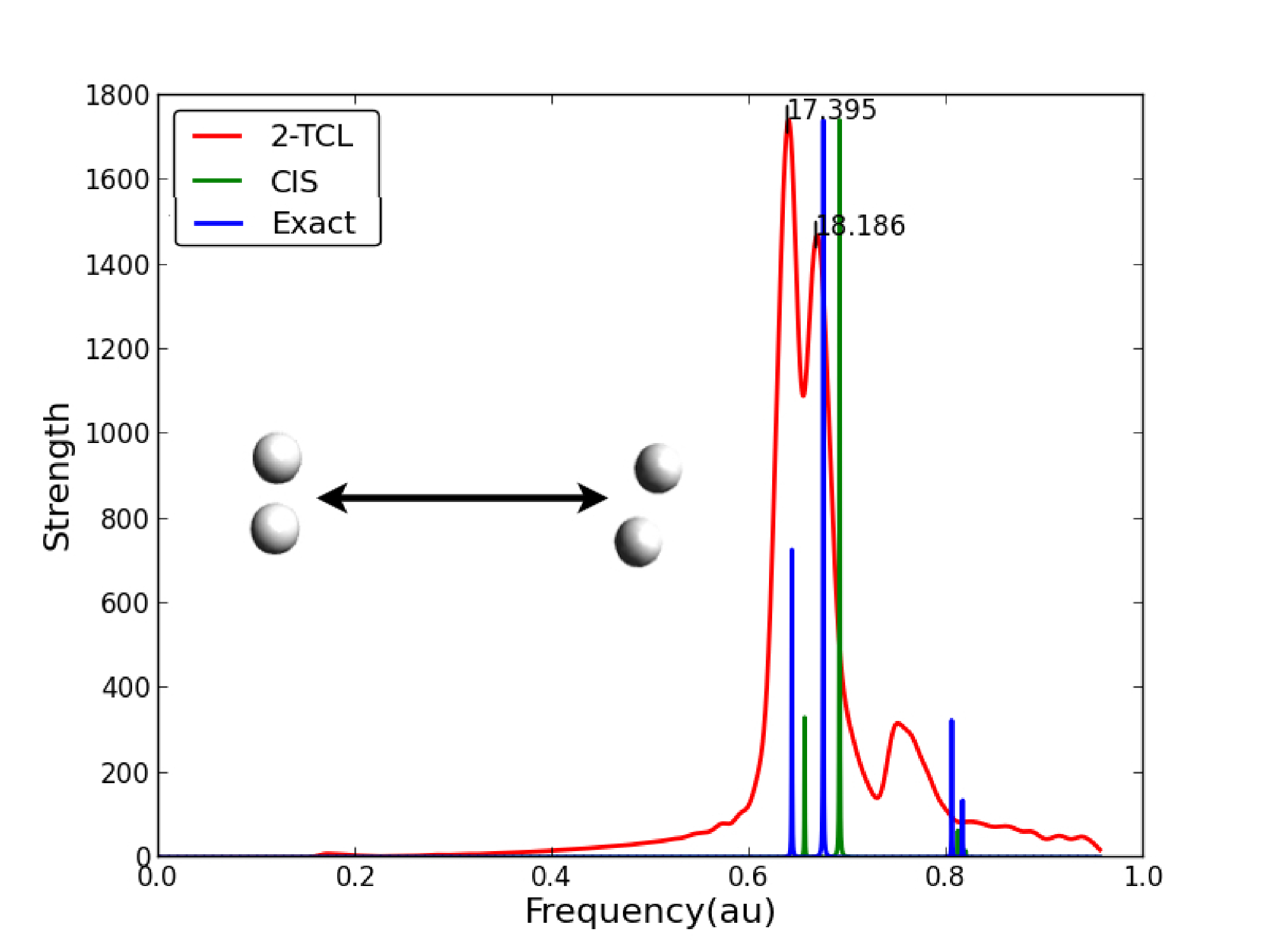}
\caption{Adiabatic linear response spectrum of a minimal model of transport between electronically excited states, two H$_2$ molecules in the DZ\cite{Dunning:1970uq} basis.}
\label{fig:LinRespTrans}
\end{center}
\end{figure}
\indent 	We have used the letter $R$ to suggest the analogy between this time-independent rate tensor and that occurring in the Redfield theory\cite{Redfield:1957fv,Redfield:1965dz}, although in this theory there are 28 such fifth and sixth rank tensors. These tensors also differ because they use the transformed, rather than the bare correlation function. The value of the integral above only depends on the values of a Laplace-transformed $B(\omega) = \int_0^\infty e^{i \omega t} B(t,0)$ at the zeroth order electronic frequencies ($\Delta$).  The Markovian rate matrix can then be calculated once in sixth order time, giving effective kinetic rates which include the effects of correlation and bath coupling. These can be used to calculate dynamics at a drastically reduced fourth order cost (since the perturbative EOM then takes the form of a single matrix product with time-independent effective Hamiltonian) or diagonalized directly to obtain spectra. The frequency independent nature of this term means that no new peaks can appear due to correlation or system-bath coupling. Only damping of dynamics between correlation and bath shifted CIS-like poles leading to Lorentzian spectra can be expected within the confines of this Markovian approximation.\\
\indent 	To give an example of how this methodology could be applied to transfer of electronic energy, we examine two Hydrogen molecules separated several bond lengths\footnote{Coordinates: ((Bohr) H: -.45 0. 0.0; H: .45 0. 0.0; H -0.48296291 0. 5.0; H 0.48296291 0. 5.2588190)} in a DZ basis. The basis has been expanded to allow for the induced dipole moment between the two molecules. Since these two molecules are well separated and held at nearly their Born-Oppenheimer minima there is no longer any strong correlation or spin-frustration in this system. The adiabatic linear response spectrum is shown in Fig (\ref{fig:LinRespTrans}), again showing good agreement with the stick spectra emerging from exact diagonalization as in $H_4$. The CIS states of this pair of molecules are well-localized. The splitting between the two peaks around 18eV corresponds to a weak coupling between them, and a distortion to the molecule on the right which has been imposed to lower the energy of the state localized on that side so that a bath can induce transfer of population. We initialize the pair of molecules into an even superposition of their excited states and couple a 2831cm$^{-1}$ oscillator at 273.0K to the HOMO, HOMO-1, LUMO and LUMO+1 with dimensionless $\tilde{M}$'s of (0.025, 0.05, .165, and .055) respectively. Over the course of $\sim$60fs (Fig. \ref{fig:Trans}) the overlap of the time-dependent state with the higher energy state 0 has halved the overlap with the lower energy state has grown, and the overall norm of the state has decreased by about 10 percent, corresponding to non-radiative decay. The rapid beating between the states corresponds to what would tight binding-model would describe as the coherence between collective states 0 and 1. The evolution of this "coherence" isn't captured well in the Tamm-Dancoff approximation because we have truncated the blocks of the propagator coupling particle-hole excitations and their conjugate hole-particle modes. 

\subsection{Untransformed and uncorrelated version}
\begin{figure}
\begin{center}
\includegraphics[totalheight=2.1in]{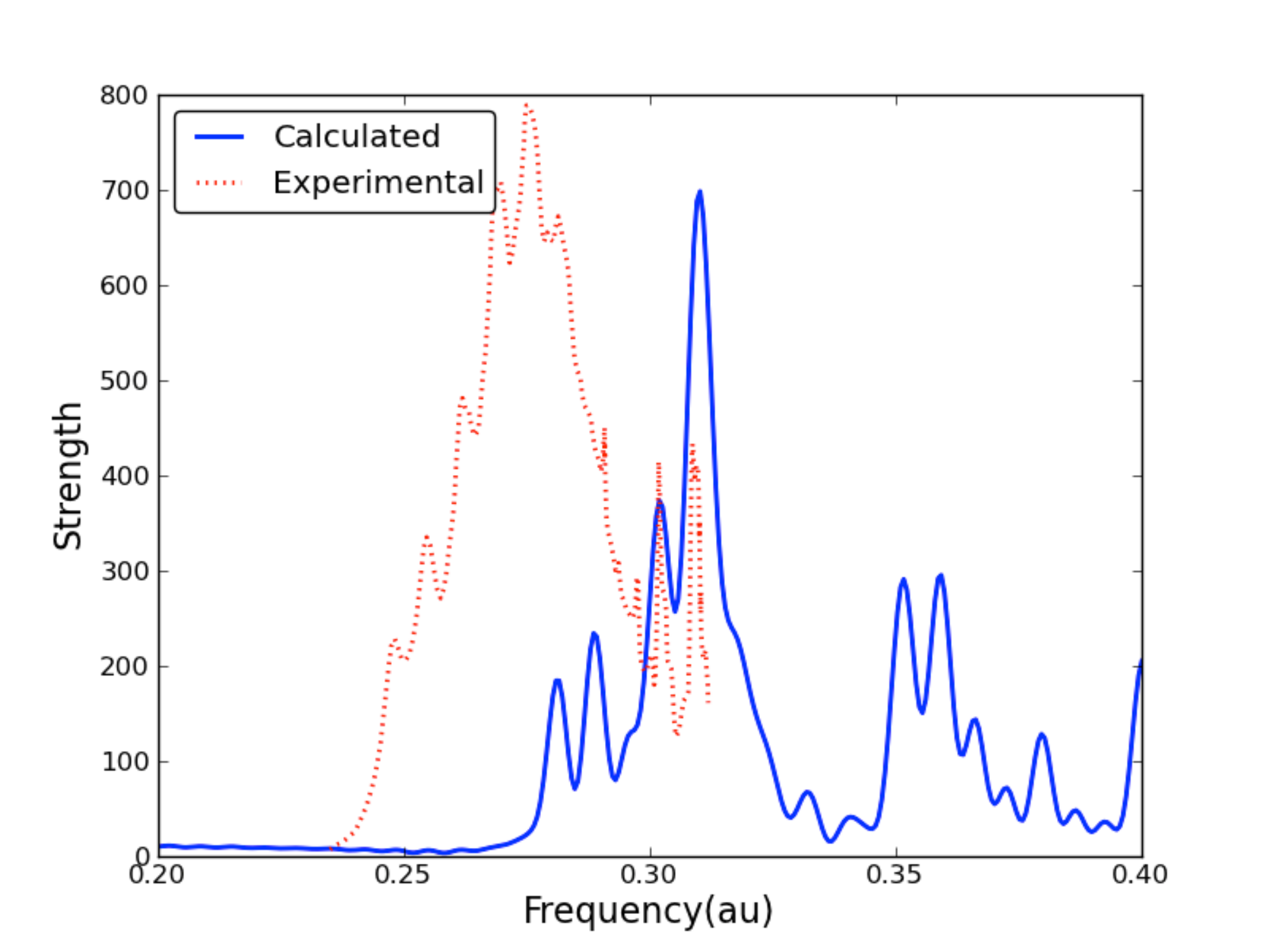}
\caption{ Experimental and simulated UV absorption cross-section of 1,1-diflouroethlyene in the region dominated by the $\pi\rightarrow\pi^{*}$ transition produced by 1500a.u. of propagation at 303K. The untransformed version of the theory is employed and correlation terms are neglected.}
\label{fig:c2f2}
\end{center}
\end{figure}

\indent	The transformed version of the theory has several interesting formal advantages, but to treat larger systems or off-diagonal system-bath couplings we have also presented the untransformed version of our theory which is similar to a propagation of the CIS wave-function with dissipative bath terms. Even in our relatively rudimentary code it is possible treat much larger systems with the undressed version especially if the electron correlation terms are neglected. To demonstrate the idea we have simulated the ultraviolet photoabsorption spectrum of 1,1-diflouroethylene\footnote{Geometry (Angstrom): F: 0.979002, -0.062874, -0.111271; C, 2.253554, 0.247708, -0.293548; F: 2.903831, -0.802928, -0.770347; C: 2.791571,1.423625,-0.041291;H: 3.843081,1.582482, -0.222345;H:2.177490,2.222097,0.344802, Basis: 6-31+G* on C, 3-21G on all other atoms}. We will compare the result of an untransformed calculation (Fig. \ref{fig:c2f2}) with available experimental data\cite{Limao-Vieira:2006fk} for the valence $\pi \rightarrow \pi^{*}$ transition. Bath bosons are positioned at the experimentally raman active frequencies, the C=C stretch at 0.214eV, the CF$_2$ symmetric mode at .114eV, and the CH$_2$ rocking mode at .162eV. We have also added a super-ohmic spectral (n=3) density with coupling constant $\alpha = 0.01$ and cutoff frequency of 5580 cm$^{-1}$. Because of a mixture of basis limitations and the absence of electron correlation the simulated absorption is centered an electron-volt higher in energy than the gas-phase experiment, but both transitions appear in a qualitatively similar way.
The somewhat irregular vibronic progression of the experimental peak hints at limitations of approximating the nuclear degrees of freedom as a harmonic bath. This suggests that in future work it may be interesting to explore more accurate ways to calculate $\langle b^\dagger(t) b(0)\rangle$, for example by propagating frozen gaussians or pursuing an Ehrenfest-type\cite{Subotnik:2010fk} scheme. In this untransformed version semiclassical or wave-packet\cite{Heller:1995uq} approaches to calculate $\langle b^\dagger(t) b(0)\rangle$ are easily combined into the electronic equation of motion.

\section{Conclusion}
\indent 	In this paper we have presented a model of correlated electronic dynamics that is transformed by a Bosonic bath. This allows us to study the effects of environmental dephasing on electronic excited states without assuming a reduced model for the electrons and their coupling to each other. Intriguingly, factors appear in the theory which damp high-rank operator expectation values. The possibility of exploiting this damping to reduce the cost of high-level electronic structure is interesting\cite{Lowdin:1994fk}. The adiabatic limit of the propagation developed here offers a useful correction to the placement of electronic energies. The linear-response spectrum of the complete model has been shown to exhibit vibronic structure. A Markovian, Tamm-Dancoff approximation to an energy transport problem has been examined, leading to bath-induced transfer of electronic energy between $H_2$ molecules, although coherence decay cannot yet be captured because several off-diagonal blocks of the transition density matrix have been neglected in the EOM of this initial work. The untransformed version of this formalism could be made efficient enough to study non-linear optical experiments\cite{Collini:2009bh,Hayes:2011zr,Tekavec:2007ly} via propagation. \\
\indent 	Accurate protocols for providing a per-orbital spectral density must be developed and the accuracy of the resulting lineshapes and lifetimes must be assessed. Simply projecting the orbital energy gradient onto the normal modes of the molecule is an obvious choice, but it would be appealing to treat the influence of surrounding molecules in a similar way. The Tamm-Dancoff equation of motion provided in this work propagates only one-block of a much larger transition density matrix which must be treated to capture coherence phenomena between particles amongst themselves (and holes). The treatment of a thermalized initial condition is an interesting, but computationally demanding question. The payoff for developing these additional features would be electronic excited states with more realism than can be offered in the adiabatic picture. These states would be naturally localized, with a size depending on vibronic structure and temperature and evolve and relax irreversibly. \\

\section{Appendix: Expressions for the correlation functions and factorization}
\indent 	Time-ordered harmonic correlation functions (HCF's) of all orders of the type $\langle \hat{X} \hat{X}^\dagger ... \hat{X} \hat{X}^\dagger \rangle $ were made available in Dahnovsky's pioneering work\cite{Dahnovsky:2007vn} by expanding the exponentials in $\hat{X}$ as power series in $\hat{A}_k = (\hat{b}_k e^{i \omega_k t} - \hat{b}^\dagger_k e^{-i \omega_k t})$ and applying Wick's theorem to the resulting $\hat{A}$ operator strings paying careful attention to the combinatorial statistics. Since we employ a master equation theory rather than a Green's function theory, our version of the HCF depends on the sign of $(t-s)$ but is otherwise the same after making the simplifications which appear in our second order theory. 
\begin{align}
B^{p_1 p_2 p_3 p_4}_{q_1 q_2 q_3 q_4}(s,t) = \langle \hat{X}^\dagger_{p_1}(t) \hat{X}^\dagger_{p_2}(t) \hat{X}_{p_3}(t) \hat{X}_{p_4}(t) 
 \hat{X}^\dagger_{q_1}(s) \hat{X}^\dagger_{q_2}(s) \hat{X}_{q_3}(s) \hat{X}_{q_4}(s) \rangle  = \notag \\
\text{exp} \{ - \sum_m \frac{1}{2}\text{Coth}(\beta \omega_m/2) (\tilde{M}^{ p_1 p_2 q_1 q_2}_{p_3 p_4 q_3 q_4}(m))^2 \} \cdot 
\text{exp} \{ - \sum_m (\tilde{M}^{ p_1 p_2}_{p_3 p_4}(m)\tilde{M}^{q_1 q_2}_{q_3 q_4}(m)) F_m(t-s) \} \notag \\ 
\text{ where: } F_m(t) = \text{Coth}(\beta \omega_m/2) \text{Cos}(\omega_m t) - i \text{Sin}(\omega_m t)
\label{eq:bcf}
\end{align}
We use an abbreviated notation (the same as $\Delta$), $\tilde{M}^{ab...}_{ij...}(s) = (\tilde{M}_s^{a} +\tilde{M}_s^{b}...-\tilde{M}_s^{i} -\tilde{M}_s^{j} ...)$. The correlation function above includes equilibrium value of the HCF. The qualitative behavior of the real part of bath integrals in terms like \ref{eq:TermBad} governs relaxation and is worth comment. If $\omega_s\sim\Delta>0$ at time $s$, and $\tilde{M}_\text{at t}\tilde{M}_\text{at s}<0$ then summed over all terms $\text{Re}(B(t))$ is negative(causing relaxation) for a time on the order of $1/(\omega_s - \Delta_s)$, after which it oscillates. If  $\Delta<0$ then relaxation occurs if $\tilde{M}_\text{at t}\tilde{M}_\text{at s}>0$. In most applications of Master equations a single, positive, spectral density is assumed for all states which basically parameterizes $\tilde{M}_\text{at t}\tilde{M}_\text{at s}$ as a function of $\omega$. This approximation is more than a mere convenience; if $\tilde{M}$ is assigned generally and different states are allowed to couple to a single frequency with different strengths it's quite easy to for some elements of the EOM to have $\tilde{M}_\text{at t}\tilde{M}_\text{at s}>0$ causing exponential growth in the Markovian limit and at short times.\\
\indent 	To treat a continuous number of bath oscillators one can introduce a continuous parameterization of $\Tilde{M}$ called the spectral density, $J_i(\omega)$, $\tilde{M}^i_{\omega_{\alpha}} = \int_0^\infty \sqrt{J_i(\omega)}/\omega \delta(\omega-\omega_\alpha)   d\omega$. The renormalization of the electronic integrals is clear given this form. With a relatively simple functional form for $J$, such as a super ohmic spectral density with cutoff parameter $\omega_c$, $J_i(\omega) = \frac{\eta_i}{6} \frac{\omega^3}{\omega_c^2} e^{-\omega/\omega_c}$, the time dependence of of correlation function $B(t,s)$ can be analytically calculated\cite{Jang:2002qf} to a good approximation. So long as $\omega_c$ is the same for all single-electron states, and only $\eta_i$ changes the whole formalism works identically with $\sqrt{\eta_i}$ taking the role of $\tilde{M}^i_{\omega_\alpha}$. We note that the requirement that the correlation function be easily integrable is only a pre-requisite for the polaron transformation. If only the time-convolutionless equation of motion is used, any correlation function which is known can be easily incorporated.\\
\indent 	Without further approximation a sixth-order number of B's must be calculated and integrated (in our code a third order Gaussian Quadrature with the electronic time-step was found sufficient), since at least two indices are shared between each $\tilde{V}$. Consider the cost limiting term Eq. (\ref{eq:TermBad}). It would be advantageous to evaluate the implied sum over $c$ and make a 5-index intermediate, since then the remaining indices are also a fifth order loop, unfortunately the HCF depends on index c. On the other hand, \emph{High-rank operator strings are \textbf{exponentially} damped by the presence of this HCF} in the Markovian limit. It seems likely that this feature could be used as a new locality principle which would lift the curse of dimensionality in the strong bath regime.\\

\acknowledgements{We would like to thank Shervin Fatehi and Jarrod McClean for help with the manuscript, and Prof. Seogjoo Jang for valuable conversations. We acknowledge the financial support of Defense Advanced Research Projects Agency grant N66001-10-1-4063, A.A.G. thanks the Camille and Henry Dreyfus and Sloan foundations, and RLE Center for EXcitonics for their generous support.}

%

\end{document}